\documentclass[12pt]{article}

\usepackage{newtxtext,newtxmath,ulem}

\usepackage{graphicx}

\usepackage[letterpaper,margin=1in]{geometry}

\renewenvironment{abstract}
	{\quotation}
	{\endquotation}

\date{\today}

\makeatletter
\renewcommand{\fnum@figure}{\textbf{Figure \thefigure}}
\renewcommand{\fnum@table}{\textbf{Table \thetable}}
\makeatother

\usepackage{scicite}

\usepackage{url}

\usepackage{xcolor}

\def\scititle{100 years of spin: fundamental physics, dark matter, exotic interactions, and all that
}
\title{\bfseries \boldmath \scititle}

\author{
	Dmitry~Budker$^{1,2,3,4\ast}$,
	Tim~Chupp$^{2,5}$,
	Klaus~Kirch$^{6,7}$,
    W.~Mike~Snow$^{8}$\and
	\small$^{1}$Johannes Gutenberg-Universit{\"a}t Mainz, 55128 Mainz, Germany.\and
	\small$^{2}$Helmholtz Institute Mainz, 55099 Mainz, Germany.\and
    \small$^{3}$GSI Helmholtzzentrum für Schwerionenforschung GmbH, 64291 Darmstadt, Germany.\and
     \small$^{4}$Department of Physics, University of California, Berkeley, California 94720, USA.\and
    \small$^{5}$Department of Physics, University of Michigan, Ann Arbor, Michigan 48104, USA.\and
    \small$^{6}$PSI Center for Neutron and Muon Sciences, 5232 Villigen PSI, Switzerland.\and
    \small$^{7}$Institute for Particle Physics and Astrophysics, ETH Zurich, 8093 Zurich, Switzerland.\and
    \small$^{8}$Department of Physics, Indiana University, Bloomigton, Indiana 47408, USA.\and 
	\small$^\ast$Corresponding author. Email: budker@uni-mainz.de\and
}

\begin{document} 





\maketitle
\tableofcontents

\begin{abstract} \bfseries \boldmath
For a century, spin has been an indispensable probe of the fundamental laws of nature. A reflection on the role of spin in shaping modern physics is presented, from the early days of quantum mechanics to the latest precision tests of the Standard Model. The significance of magnetic and electric dipole moments in testing $CP$ and $CPT$ symmetries is surveyed, along with the ongoing searches for exotic spin-dependent interactions that may reveal the nature of dark matter and its connection to spacetime geometry. Through these vignettes, it is shown that spin continues to provide a fresh perspective on the most profound questions in physics today.
\end{abstract}

\noindent


\section{Introduction}
\subsection{The goals of this paper}
The authors of this article are faced with a nearly impossible task: in a few pages, we need to reflect on the connection between spin and fundamental physics. The enormity of this challenge comes from the fact that our modern understanding of nature is impossible to disentangle from spin physics and that such cornerstones of physics as quantum mechanics, quantum field theory, and fundamental symmetries are all intrinsically related to spin. It is for this reason that we decided to present a collection of spin-related ``vignettes'' which, without any attempt at completeness, will hopefully give the reader a fresh perspective on the role and importance of spin.\footnote{Various complementary aspects of spin physics are discussed by other authors in this collection, for instance, the role of spin in quantum chromodynamics is covered in a review by A.\,Deshpande1, D.\,E.\,Kharzeev, and J.\,Liao.}

A key point is that, starting with the early years of quantum mechanics, spin has always played a central role in establishing our fundamental physical picture of the world. This is continuing today, as we illustrate with examples in this article.

\subsection{Discovery of spin}

Just over 100 years ago, M.\,Catal\'{a}n~\cite{Catalan1922} and A.\,Land\'{e}~\cite{Lande1923_Pt1,Lande1923_Pt2} independently pointed out a puzzling splitting of atomic spectral lines in a magnetic field. An explanation was proposed in 1924 by W.\,Pauli, who introduced the concept of ``two-valuedness'' inherent to electrons~\cite{Pauli1925_ExclusionPrinciple}. This idea was further fleshed out by G.\,Uhlenbeck and S.\,Goudsmit in the following year, when they introduced the notion of electron spin~\cite{Uhlenbeck1925_SpinHypothesis}. Their bold hypothesis, though initially met with skepticism, would go on to revolutionize and shape our understanding of quantum mechanics. The discovery of spin not only resolved long-standing puzzles of the time but also opened doors to new technologies, from magnetic resonance imaging\footnote{See the articles on magnetic resonance imaging in this collection.} to spin-based computing \cite{Kurebayashi2026_spin_computing}.

\subsection{Proton spin}
\label{subsecProtonSpin}

Proton spin was introduced by David Dennison in a 1927 paper to explain the specific heat of hydrogen at low temperature, which showed a sharp drop below about 200\,K.  
The discrepancy was resolved by introducing spin 1/2 for the protons and invoking the consequence of the Pauli principle for H$_2$. For H$_2$, there are two possible proton-spin arrangements: total spin $S=0$ parahydrogen with proton spins opposite and $S=1$, orthohydrogen, with one and three states, respectively. The wavefunction of H$_2$ is overall symmetric, which is consistent with a rotational ground state $J=0$ with total electron spin zero and total proton spin zero, {\i.e.} parahydrogen. The energy required for a rotational excitation and transition from orthohydrogen to parahydrogen corresponds to about 200\,K. Below that temperature, the specific heat drops significantly \cite{Dennison1927_SpecificHeat}.  Note that, in contrast, deuterium, nuclei are spin-1 bosons, and symmetry requires symmetric nuclear spin wave function, so the $J=0$ rotational ground state is ``orthodeuterium''.


\section{Spin has its moments: magnetic and electric}

\subsection{Magnetic and electric dipole moments in the Dirac equation}
\label{subsec:Dirac}

Paul Dirac is well known for his work combining quantum mechanics and relativity, resulting in a theoretical prediction of the electron $g$-factor and the interpretation of negative-energy states as antimatter.
The equation which now bears Dirac's name was also worked out independently at about the same time by Hendrik Kramers: although his derivation was not published for many years~\cite{Kramers1935} after Dirac's work appeared~\cite{Dirac1927}, Kramer's work was well known to his contemporaries~\cite{Dresden1987, TerHaar1998}. 

In Dirac's 1927 paper ``The Quantum Theory of the Electron", the Hamiltonian for an electron with spin 1/2 in magnetic and electric fields was derived from a scalar potential $A_0$ and a vector potential $\vec{A}$, leading to two terms (note the non-SI, specifically cgs, units):
\begin{equation}
    \frac{e\hbar}{2mc}(\vec{\sigma}\cdot \vec{H}) +\frac{ie\hbar}{2mc}\rho_1(\vec{\sigma}\cdot \vec{ E})\,,
\end{equation}
where $\vec{\sigma}$ are $4\times 4$ spinor matrices. The first term, the interaction with a magnetic field $\vec{H}$ with the magnetic moment $|\vec\mu|=g\frac{e}{2mc}\frac{\hbar}{2}$, with $g=2$, agrees with the spinning-electron model of Goudsmit and Uhlenbeck.  The second term, the interaction with an electric field $\vec{E}$, includes  $i=\sqrt{-1}$. This led Dirac to doubt that it had any physical meaning since the electric field is real. In fact, we now understand that this imaginary term changes sign under the time-reversal transformation, and the $(\vec{\sigma}\cdot \vec{ E})$ is also odd under parity transformations. The ``electric moment'' is therefore T and P odd, and due to CPT invariance, it is also CP violating. We further discuss electric dipole moments (EDMs) in Sec.\,\ref{sec:EDM}.



\subsection{Magnetic moments and spin}
Dirac revealed that the magnetic moment of the electron with spin $\hbar/2$ should be proportional to a factor $g=2$. Measurements of zero-field splittings in hydrogen and deuterium by Rabi et al.\cite{NafeNelsonRabi1947_HDhfsandelectongm2,NafeNelson1948_HDhfsandgm2} hinted  that $g\ne2$. RF spectra of excited-state $n=2$ hydrogen by Lamb and Retherford were also inconsistent with calculations based on Dirac theory~\cite{LambRetherford1947_LambShift}. Precise measurements of the hyperfine spectra of Ga and Na by Kush and Folley~\cite{KushandFolley1948_electronmu} showed results consistent with $g_l=1$ and $g=2\cdot(1.00119)$ with uncertainties of 50 parts-per-million (ppm). The corrections to the electron $g$-factor were explained by quantum-field theory of electromagnetic interactions or quantum electrodynamics (QED), developed by Schwinger, Feynman, Tomonaga, and Dyson followed by many others~\cite{Dyson1947_QEDUnified}. 

Over decades, the magnetic moment anomaly, generally expressed as $a=(g-2)/2$ or just $g-2$, of the electron and muon have been calculated with increasing precision. The largest corrections, due to QED, have been calculated to 5th order in the fine-structure constant $\alpha$~\cite{Aoyama2018_10thorderQEDgm2}. Electroweak (EW) corrections due to the exchange of massive vector bosons for the electron and muon contribute to $g-2$ at the 1.3\,ppm level, and have been calculated with a precision of less than 1\% or 10 ppb~\cite{Czarnecki2003_EWgm2}. Both QED and EW correction calculations are perturbative, that is the expansion to higher order converges. In contrast, strong interaction contributions, specifically hadronic vacuum polarization loops with $\pi^+$-$\pi^-$ have presented the greatest challenge and largest uncertainties. Note that the vacuum polarization loop contributions from heavier particle-pairs ($\mu^+-\mu^-$ and $\pi^+$-$\pi^-$) scale as the particle mass, and the contributions as well as the uncertainties are much larger for the muon than for the electron. A dispersion-relation approach based on the optical theorem that relates the cross section for $e^+$-$e^-$ $\rightarrow$ $\pi^+$-$\pi^-$ was considered the most reliable calculation until recently~\cite{Aoyama2020_muongm2TheoryWhitePaper2020}. Lattice QCD methods have been applied to calculate hadronic vacuum polarization as well the  hadronic light-by-light contribution~\cite{Aliberti2025_muongm2TheoryWhitePaper2025}. The current muon $g-2$ theory is summarized in the lower part of Fig.\,\ref{fig:Muongm2History} and discussed further in  Sec.\,\ref{subsec:muon_g-2}. 

Hadronic vacuum-polarization contributions dominate the uncertainty of the Standard-Model calculation of the muon magnetic-moment anomaly, but are currently small compared to the experimental precision on the electron's anomaly, but may soon become significant with anticipated improved measurement of the electron anomaly.

Precise measurements of the magnetic moment of the electron and later the muon became the testing ground of the power of quantum-field theory to precisely calculate the interaction with the quantum vacuum. These interactions encompass all of the standard model particles and interactions as well as potential Beyond-Standard-Model (BSM) interactions, and tensions between measurements and Standard-Model calculations have motivated  recent and prospective experimental efforts discussed below.

\begin{figure} 
	\centering
	\includegraphics[width=0.49\textwidth]{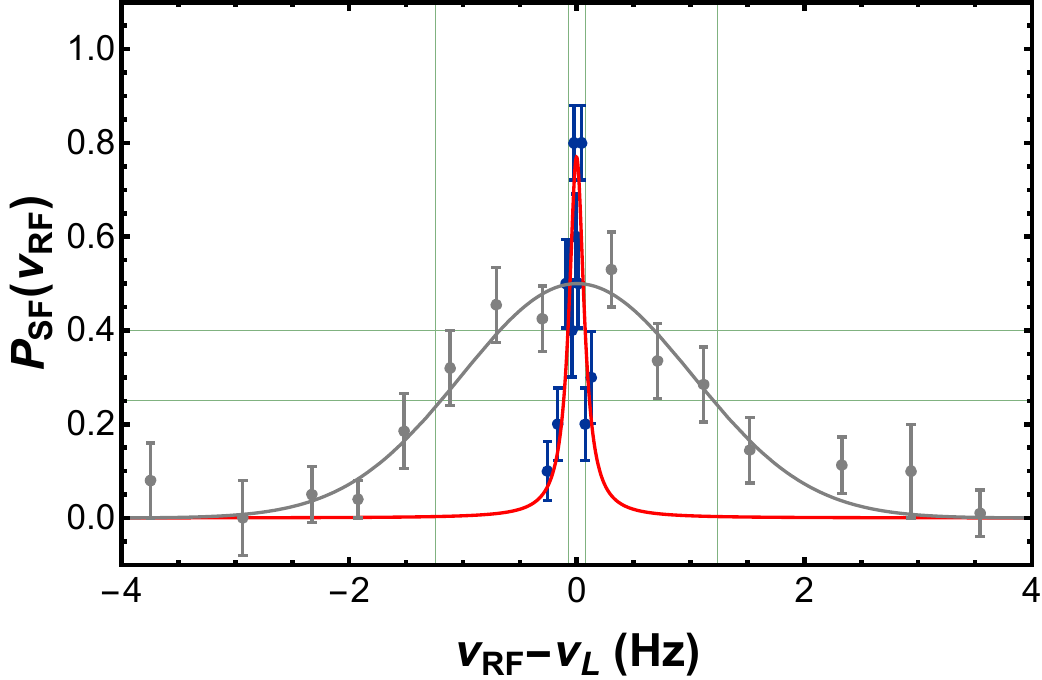}
    	\includegraphics[width=0.49\textwidth]{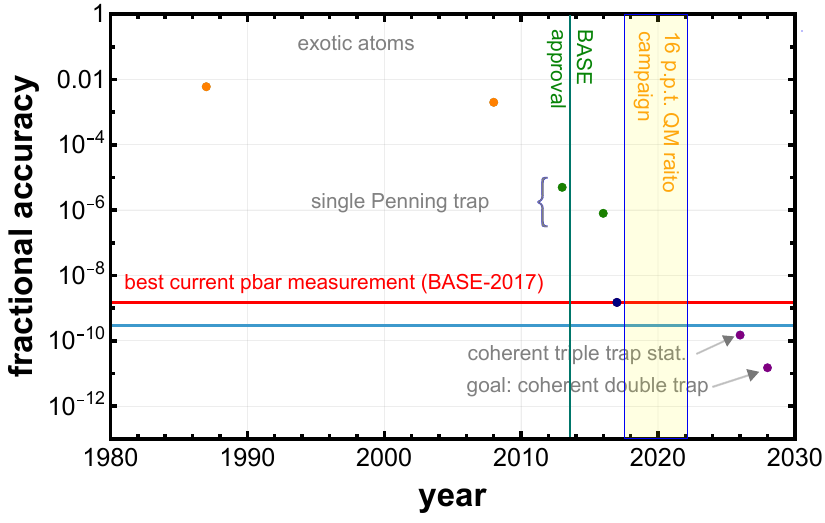} 
	\caption{\textbf{Magnetic resonance of a single antiproton spin \cite{Latacz2026_BASE_Nature} and fractional resolution achieved in measurements of the antiproton magnetic moment as a function of time.}
Above: The plot shows the most recent results from the BASE collaboration (resonance width of 0.156(4)\,Hz FWHM, blue points) and the least-squares fit of a Voigt profile to the data (red line). The gray data points and fit are from the previously measured antiproton resonance with a width of 2.5(2)\,Hz FWHM \cite{Smorra2017_BASE_Nature}. Below: A fractional accuracy on the $10^{-3}$ level was reached with exotic-atom spectroscopy reached (orange); measurements in single Penning traps achieved resolutions at the ppm level. Upon approval, the BASE collaboration measured the antiproton magnetic moment with a frational resolution at the $10^{-9}$ level. It is expected that recently introduced coherent spectroscopy will allow for triple-trap measurements at the level of $\approx0.1\,$ppb. BASE is currently developing a single-particle double-trap technique with cyclotron frequency measurements based on phase sensitive detection, with projected fractional resolution $<20\,$ppt. Figure courtesy S.\,Ulmer (BASE).
		}
	\label{fig:PBAR_Resonance} 
\end{figure}

Symmetry under CPT, the combined discrete transformations of charge conjugation (C), parity (P), and time-reversal (T) is a fundamental principle underlying any quantum field theory \cite{streater1978pct}. CPT symmetry requires that the magnetic moment of a particle  and its antiparticle  are equal in magnitude but opposite in direction with respect to the angular momentum $\vec J$, {\it i.e.} with $\vec\mu=\mu\vec J$, $\bar\mu\langle\vec J\rangle=-\bar\mu\langle\vec J\rangle$.
Measurements of magnetic dipole moments of the electron and positron, negative and positive muons, and proton and antiproton (see Fig.\,\ref{fig:PBAR_Resonance}) all confirm CPT symmetry. Measurement of the proton and antiproton EDMs (see Sec.\,\ref{subsec:EDM searches}), in principle possible in the same storage ring \cite{alexander2025statusprotonedmexperiment}, would provide a CPT test of a CP-violating observable.

\subsection{Direct measurements of the electron $g$-factor}
By the early 1950s experiments with electrons bound in atoms had established and QED calculated corrections to the the electron magnetic moment $g$-factor. 
Measurement of  unbound (free) electrons, for example using the Stern-Gerlach approach were considered with a long debate about the limitations imposed by the uncertainty principle for position and momentum of the electron, see, for example, \cite{MottandMasey1949_ThAtomCollisions}. H.\,R.\,Crane and students pioneered adapting Rabi's magnetic-resonance method to an electron-beam Mott double scattering for electron polarimetry\cite{LouisellPiddCrane1954_electrongm2}. Their subsequent experiments employed a magnetic bottle to essentially trap the electrons for up to about 2\,ms with the most precise result for the electron  $g$-factor $g_e$ 
with an uncertainty of 27\,ppb (parts per billion) on $g_e$ and 23\,ppm on $a_e$ \cite{WilkinsonCrane1963_electrongm2}.

Dehmelt and collaborators developed measurements with a single or a few electrons in a Penning trap, which used a magnetic field to radially confine the electrons in cyclotron orbits and a quadrupole electric field to confine the electrons along the magnetic-field axis. This bound system of electrons in electromagnetic fields was called ``geonium''\cite{Van_Dyck1976_geonium}. The electron motion was detected by measuring the image-charge current induced on the electrodes, which also served to damp the motion or cool the energy of the electron motion by coupling the current to a cooled resistor. Several stages of improvement provided the result 
with an uncertainty of 4\,ppt (parts per trillion) on $g_e$ and 2.7\,ppb on $a_e$~\cite{vanDyck1987_electronpositrongm2}. 

Gabrielse and collaborators undertook a series of increasingly precise measurements with a single electron in a cylindrical Penning trap~\cite{Odom2006_Electrongm2,Hanneke2008_Electrongm2,Fan2023_Electrongm2}. Improvements over two decades include inhibiting spontaneous emission and cooling the electron to the cyclotron quantum ground state. The currently most precise result is 0.11\,ppb uncertainty on $a_e$ or 0.13\,ppt on $g_e$~\cite{ Fan2023_Electrongm2}. 

The Standard-Model calculation of $g_e$ uses the measured fine-structure constant $\alpha$ as an input. Currently, $\alpha$ is most precisely determined from the recoil energy from photon emission by  cesium~\cite{Parker2018_alphaCs} or rubidium~\cite{Morel2020_AlphaRb} atoms. The precision on $\alpha$ from these measurements is 200\,ppt and 81\,ppt, respectively, but they disagree by about 5.5$\sigma$. Splitting the difference, this can be incorporated into agreement of the measurement and Standard-Model calculation with an uncertainty of 0.7\,ppt. This agreement can be interpreted as constraining the electron charge radius to less than about $3\times 10^{-19}$\,m and setting limits on the scale of electron substructure~\cite{ Fan2023_Electrongm2}. Alternatively, measurement of $g_e$ can be used to determine $\alpha$, which would fall between the cesium and rubidium values with an uncertainty of 111 ppt. 

Improvements to the experimental determination of $g_e$ may soon reach a level at which the hadronic contributions for the electron are significant. This would introduce a new challenge to the Standard Model calculations to complement the muon $g$-factor discussed in the next section.

We cannot close this section on the free-electron $g$-factor without mentioning the tremendous  progress made in measuring bound-electron $g$-factors using  highly charged ions in Penning traps. Recent developments, for example, with H, Li, and B-like $^{118}$Sn ions~\cite{Morgner2023,Morgner2025a,Morgner2025b} provide sub-ppb measurements, challenge higher-order bound-state QED and nuclear theory, and might eventually also  lead to an independent extraction of $\alpha$.


\subsection{Muon $g$-2}
\label{subsec:muon_g-2}

Muon magnetic-moment anomaly measurements, in part motivated by historical tensions of QED and electron $g$-factor measurements began in the late 1950s at Brookhaven~\cite{Coffin1958_FirstMuongm2} and early 1960s at CERN ~\cite{Charpak1965_CERNMuongm2}. Magnetic storage-ring efforts began at CERN, and then a major effort through the 1990s was mounted at Brookhaven. A decade later, the Brookhaven magnet was rebuilt, and an experiment based on the same concepts with new instrumentation was undertaken at Fermilab. The combined precision on $g_\mu-2$, dominated by the last three years of data from Fermilab is 124\,ppb~\cite{Aguillard2025_FinalMuongm2}.

Accelerator muons resulting from pion-decay are polarized, that is the spin is strongly correlated with the momentum. The magnetic-moment anomaly $g_\mu-2$ can be measured with polarized muons injected into a magnetic storage ring, which confines the muons radially. In analogy to the Penning trap, quadrupole electric fields are required for confinement in the third dimension. For relativistic muons, the magnetic moment coupling to  magnetic ($\vec B$)  and electric ($\vec E$) fields leads to precession of the spin with respect to the momentum given by the vector anomaly frequency 
\begin{equation}
\vec\omega_a=\vec\omega_s-\vec\omega_c=a_\mu\frac{e}{m_\mu}\vec B
+\frac{e}{m_\mu}\left [\frac{a_\mu(\gamma^2-1)-1}{\gamma^2-1}\frac{\vec \beta\times \vec E}{c}\right ]+\frac{e}{m_\mu}\frac{\gamma}{\gamma+1}\vec\beta(\vec\beta\cdot\vec B)\,.
\label{eq:BMTEquation}
\end{equation}
Here $\vec\omega_s$ is the spin-precession angular frequency, $\vec\omega_c$ is the cyclotron frequency, $\gamma$ is the relativistic Lorentz factor, $\vec \beta$ is the muon velocity divided by the speed of light, and $e$ and $m_\mu$ are the charge and mass of the muon, respectively. This is the Bargmann-Michel-Telegedi equation~\cite{Bargmann1959_BMTEquation} as rewritten, for example in~\cite{Aquillard2024_Muongm2Run23PRD}.
The first term in Eq.\,\ref{eq:BMTEquation} is used to measure $ a_\mu$, and requires determining $\omega_a$ and the magnetic field averaged by the muons over space and time with absolute calibration\cite{Flay2021_Muongm2H2OCalibration,Farooq2020_Muongm23HeCalbration}.
The second term would lead to significant systematic errors due to the electric field, however the term can be approximately canceled up to small corrections for $\gamma\approx 29.3$ - corresponding to what is called the magic energy. The third term proportoinal to the muon velocity component along $\vec B$ also gives rise to small corrections. 

\begin{figure} 
	\centering
    \vskip -0 truein
	\includegraphics[width=1.0\textwidth]{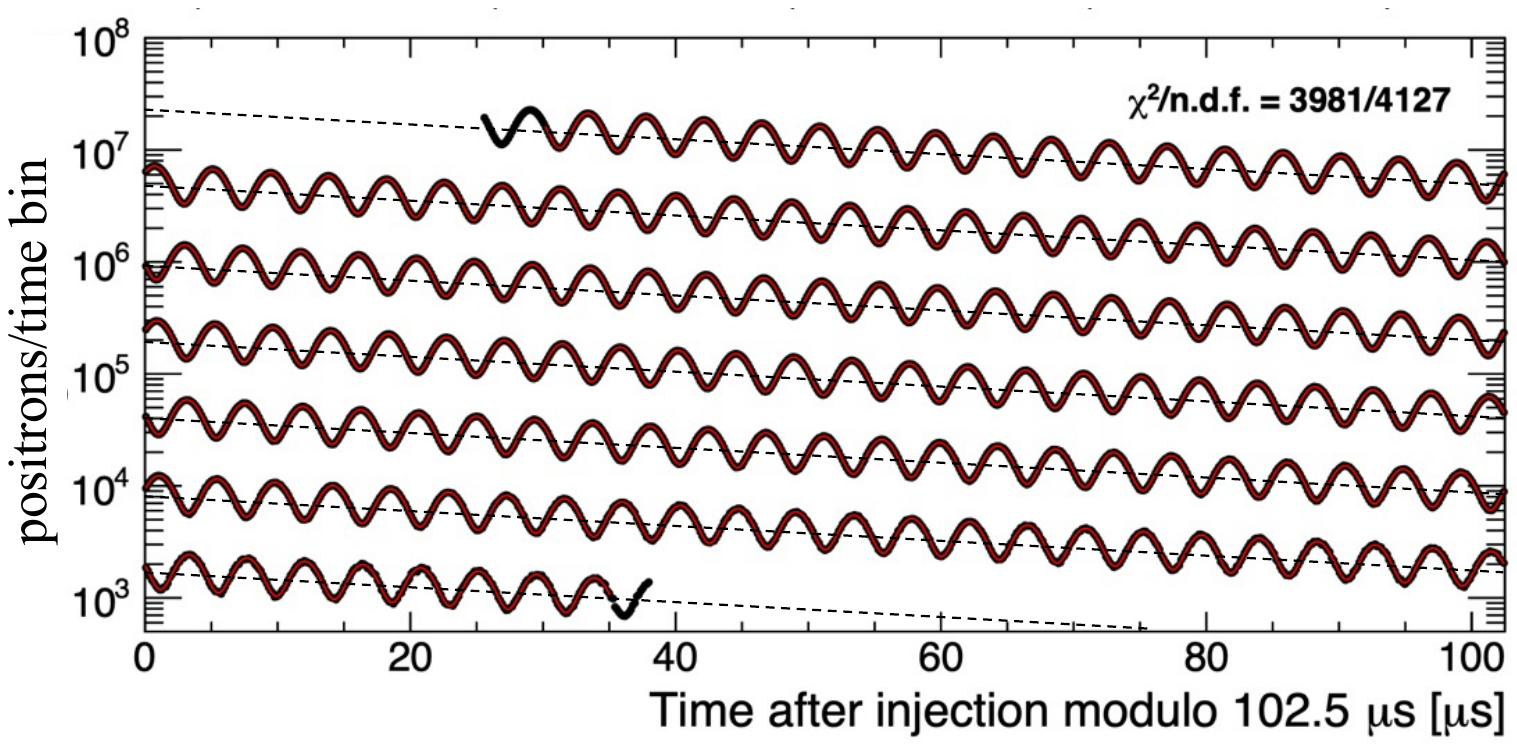} 
	\vskip -1.25 truein
\caption{\textbf{Wiggle-plot showing detected-analyzed positrons above a fixed energy threshold.} Dashed lines indicate decay of muons in the lab frame. Adapted from \cite{Aguillard2023_Muongm2Run23}. 
}
	\label{fig:gm2WigglePlot} 
\end{figure}

The energy of decay positrons (for $\mu^+$) in the lab is correlated with the relative orientation of the muon spin and momentum due to parity violation. 
Calorimeters detect the positrons above an optimum energy threshold resulting in the ``wiggle plot'' shown in Fig.\ref{fig:gm2WigglePlot}, a truly beautiful demonstration of many modern-physics concepts. For example, note that exponential decay trend of muon counts - linear on the semi-log-scale - has time constant of about 60\,$
\mu$s,   the lab-frame lifetime $\gamma\tau_{\mu}$ (with $\tau_{\mu}=2.2\,\mu$s) due to  relativistic time dilation.

\begin{figure} 
	\centering
    \vskip -0.75 truein
   \includegraphics[width=0.65\textwidth]
    {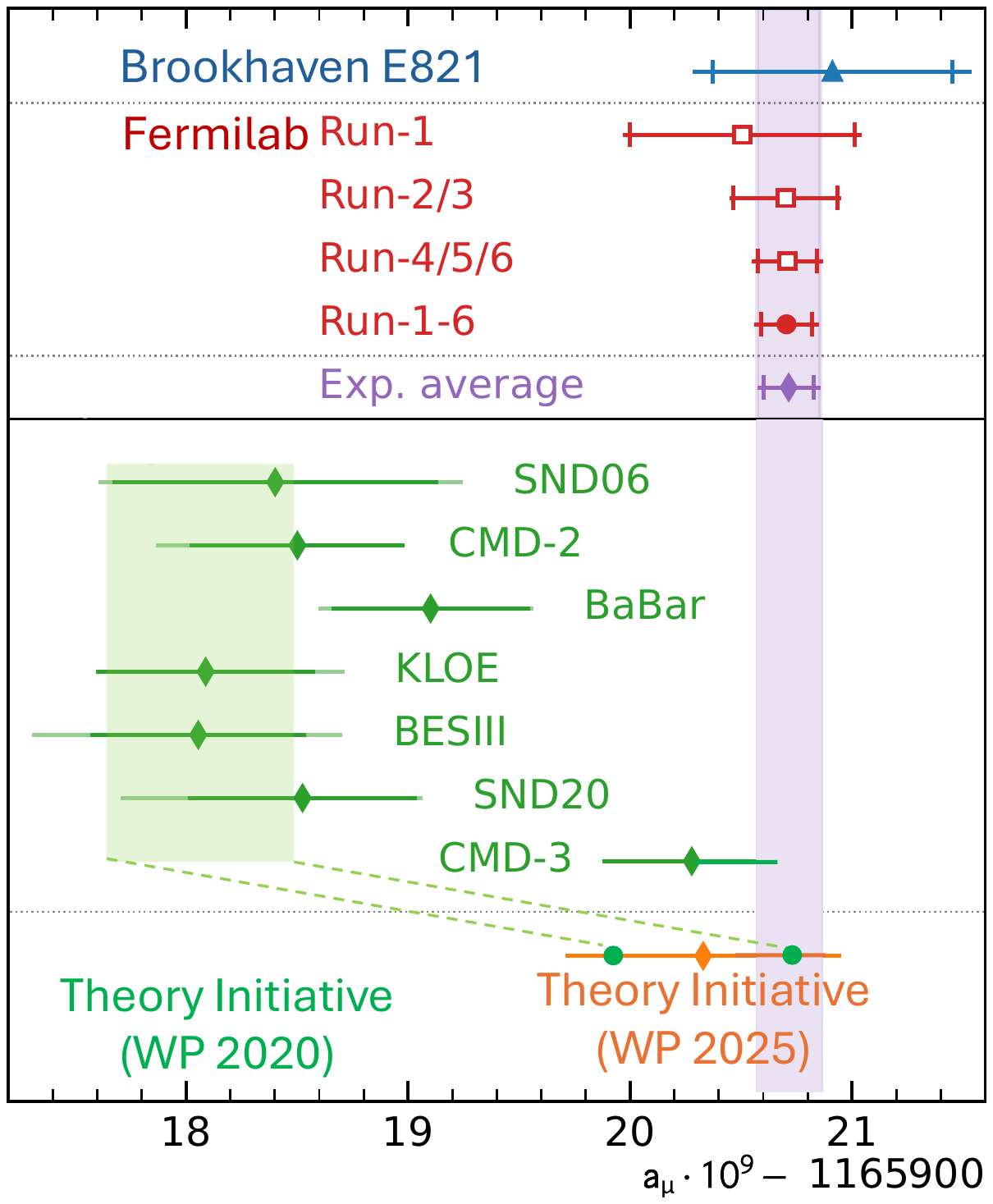} 
	\vskip -0.5 truein
    \caption{\textbf{History of  muon $g-2$.}  Measurements and Standard-Model calculations with semi-empirical (labeled WP2020) and Lattice-QCD (labeled WP2025) hadronic vacuum-polarization contributions. Adapted from \cite{Aguillard2025_FinalMuongm2}.}

	\label{fig:Muongm2History} 
\end{figure}

The history of measurements at Brookhaven and Fermilab and the Standard-Model calculations of $a_\mu$ are presented in Fig.\,\ref{fig:Muongm2History}. The experimental results are strikingly consistent as the statistics-limited uncertainties have improved by a factor of four. The Theory-Initiative estimated uncertainty of the Standard-Model calculation has increased to account for disparate semi-empirical hadronic-vacuum polarization results with recent CMD3 results and the tension of the semi-empirical and Lattice QCD calculations, see ~\cite{Aliberti2025_muongm2TheoryWhitePaper2025}. We anticipate that these estimated uncertainties will improve over time.

The magic-energy storage ring approach was statistics limited, and the recently commissioned PIP-II linac at Fermilab and improved storage-ring efficiency could conceivably provide a factor of 10 greater detected muon decay rates. A significant improvement would also require reducing the dominant systematic effects, roughly balanced between the magnetic field (mapping, tracking and calibration) and beam-dynamics effects, partly from the second and third terms in Eq.\,\ref{eq:BMTEquation}. 

A new approach to measure muon $g-2$ experiment with muonium (the $\mu^+$-$e^-$ atom) source is planned at JPARC~\cite{Abe2019_JPARCMuongm2}. Although precision of about 450\,pbb, similar to the combined Brookhaven results, is anticipated, significantly different systematic effects are expected  with a factor of 10 lower muon momentum, much smaller emittance eliminating the need for confining electric fields, and  a factor of 20 smaller dimension magnet that is much more precise and stable. 

\section{Electric dipole moments}
\label{sec:EDM}

\subsection{EDM history}
\label{subsec:EDM history}

In 1950, Purcell and Ramsey discussed the ``Possibility of Electric Dipole Moments for Elementary Particles and Nuclei''~\cite{Purcell1950}. Mostly referring to the belief in parity conservation, they argued that `suggestive theoretical symmetry arguments' must indeed be tested experimentally. They gave a first estimate for a limiting size of the EDM of the neutron, deduced from scattering experiments investigating the neutron-electron interaction. Interestingly, they thanked their student, Mr.\,Smith, `for suggesting an important correction  to' their `original calculation on the neutron-electron interaction experiment'. 
They also mention that they were undertaking a beam experiment with Mr.\,Smith to directly measure the neutron EDM, essentially using Ramsey's novel magnetic-resonance method of separated oscillatory fields~\cite{Ramsey1949,Ramsey1950} for which he was subsequently awarded the Nobel Prize in 1989.

\begin{figure}
\includegraphics[width=1.0\textwidth]{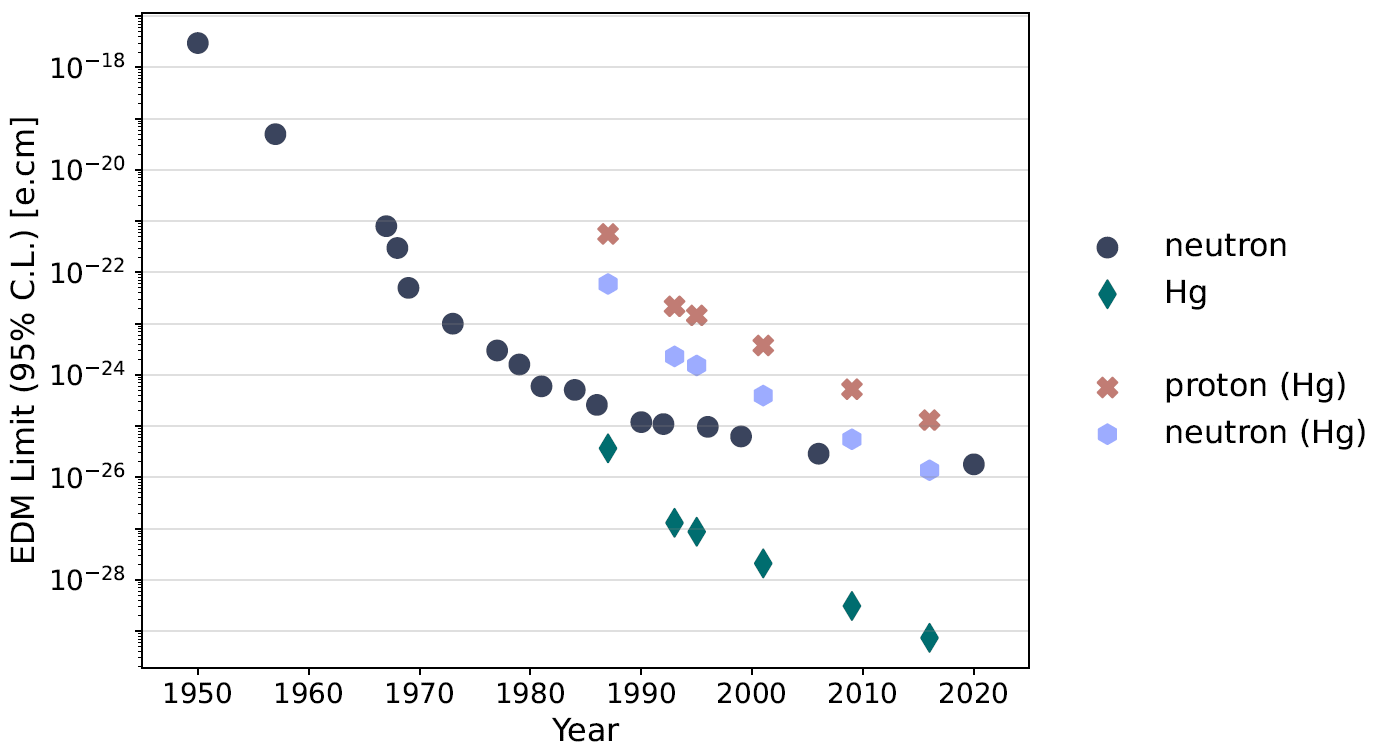}
\caption{\textbf{Nucleon EDM limits}. The history of direct limits on the neutron EDM from the first experiment~\cite{Smith1957} until today at $2\times10^{-26}e$cm~\cite{Abel2020} along with the EDM limits for the $^{199}$Hg atom, providing the most precise directly measured limit of any EDM today at $7.4\times10^{-30}e$cm~\cite{Graner2016,Graner2017}.
From the $^{199}$Hg measurement, one can with some disputed large uncertainties and some suppression factor, also derive indirect limits on the neutron and proton EDMs~\cite{Dmitriev2003}, displayed without uncertainties. The direct and indirect neutron limits are about equal, for the proton no direct limit is available yet. Figure credit: T.\,Hume, PSI.}
\label{fig:nucleonEDM}
\end{figure}

\begin{figure} 
	\centering	\includegraphics[width=1.0\textwidth]{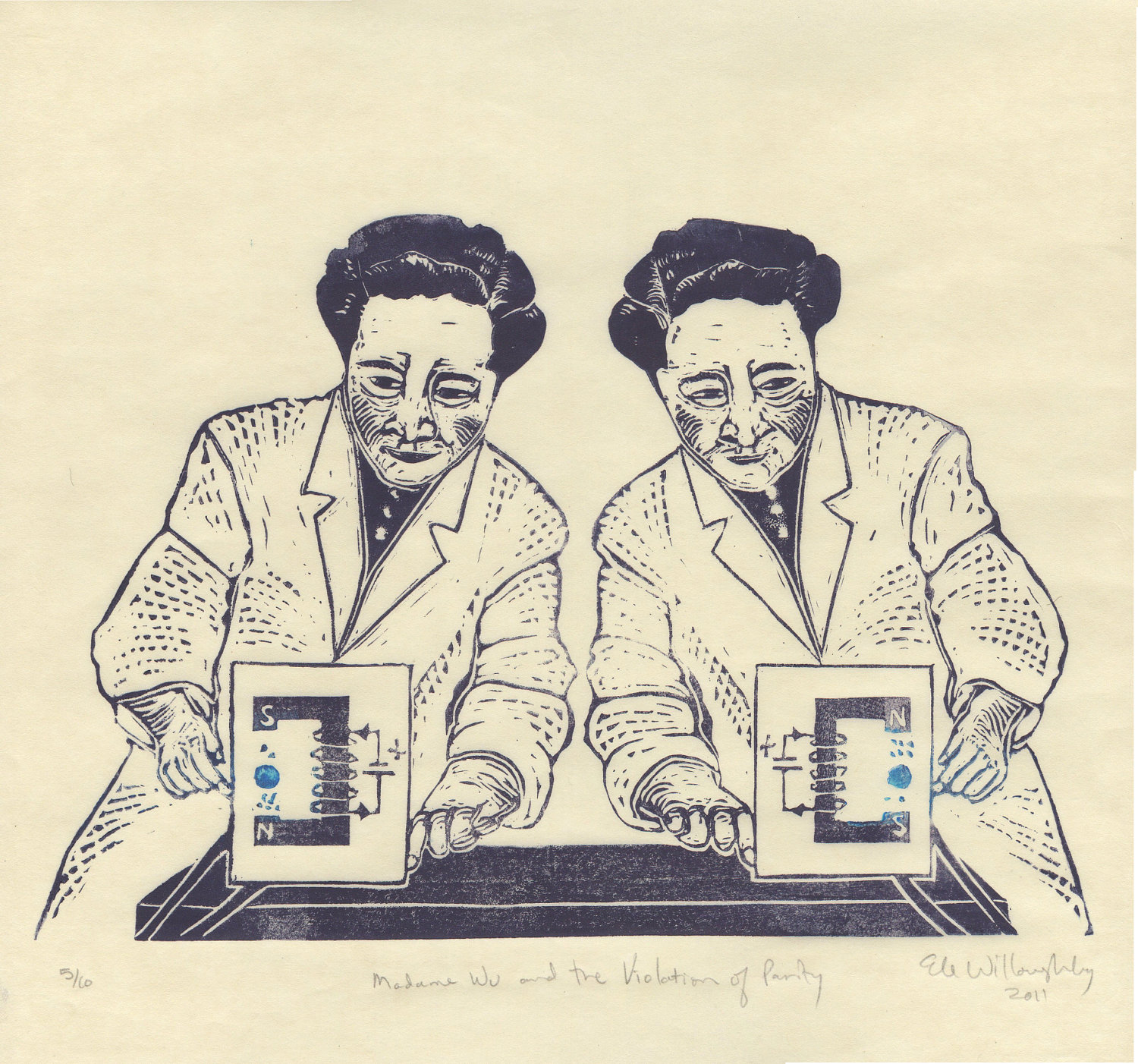} 
	\caption{\textbf{The discoverer of parity violation, C.S.\,Wu and her famous experiment \cite{Wu1957_PV}.}
The figure shows and image of a lino print by Ele Willoughby (reproduced with the artist's permission). On the left, Madame Wu is depicted with her apparatus where beta-decay of polarized $^{60}$Co nuclei (the blue ball) was studied, in which it was shown that beta-decay electrons are preferentially emitted in the direction opposite to the direction of the nuclear spin. This is significant because in a ``mirror-image'' experiment (right), one would find that beta-electrons are preferentially emitted along the spin, which is NOT observed, signifying left-right asymmetry of nature.}
	\label{fig:MadameWu} 
\end{figure}

Ramsey himself later often mentioned that although they performed the experiment in the early 1950s they were discouraged from quickly publishing by the strong belief in parity conservation. As this belief broke down with the experiments of Wu, Ambler et al. studying beta decay of spin-polarized $^{60}$Co nuclei  in 1956~\cite{Wu1957}, see Fig.\,\ref{fig:MadameWu}, and subsequent muon-decay experiments of Garwin, Lederman, Weinrich~\cite{Garwin1957} and Friedman and Telegdi~\cite{Friedman1957}, the first direct experimental limit of the neutron EDM was published by Smith, Purcell and Ramsey in 1957~\cite{Smith1957}.\footnote{Note the promotion of the PhD student from an acknowledgment to being the first author on the paper. Incidentally, J.\,H.\,Smith was later an undergraduate physics instructor of one of us (W.M.S.) at the University of Illinois.} 
With the possibility of symmetry violations, at least in weak interactions, now being beyond esoteric ideas, Ramsey also published in 1957 on the possibility of time reversal violation~\cite{Ramsey1958} and the need for experimental tests against theoretical bias. Here again, the neutron EDM was the first choice to be pursued and has been among the most sensitive probes since;\footnote{Presently, several international collaboration are working towards 1-2 orders of magnitude improvements in sensitivity of neutron EDM measurements~\cite{Athanasakiskaklamanakis2026}} see Fig.\,\ref{fig:nucleonEDM}. Interestingly, immediately after the publication of the first neutron result, still in 1958, a muon EDM measurement was carried out~\cite{Berley1958}, as in the previously mentioned parity violation studies, benefiting from polarized muons obtained in pion decay and polarization analysis due to the muon decay asymmetry. Already mentioned in~\cite{Berley1958} from a private communication with G.\,Feinberg is an indirect limit of order 10$^{-13}e$cm for the electron, estimated from the Lamb shift in hydrogen \cite{Feinberg1958}. The work by Feinberg~ was submitted on the same day as the muon result, but it was published later, back to back with a similar result from Salpeter~\cite{Salpeter1958}.

\begin{figure}
\includegraphics[width=1.0\textwidth]{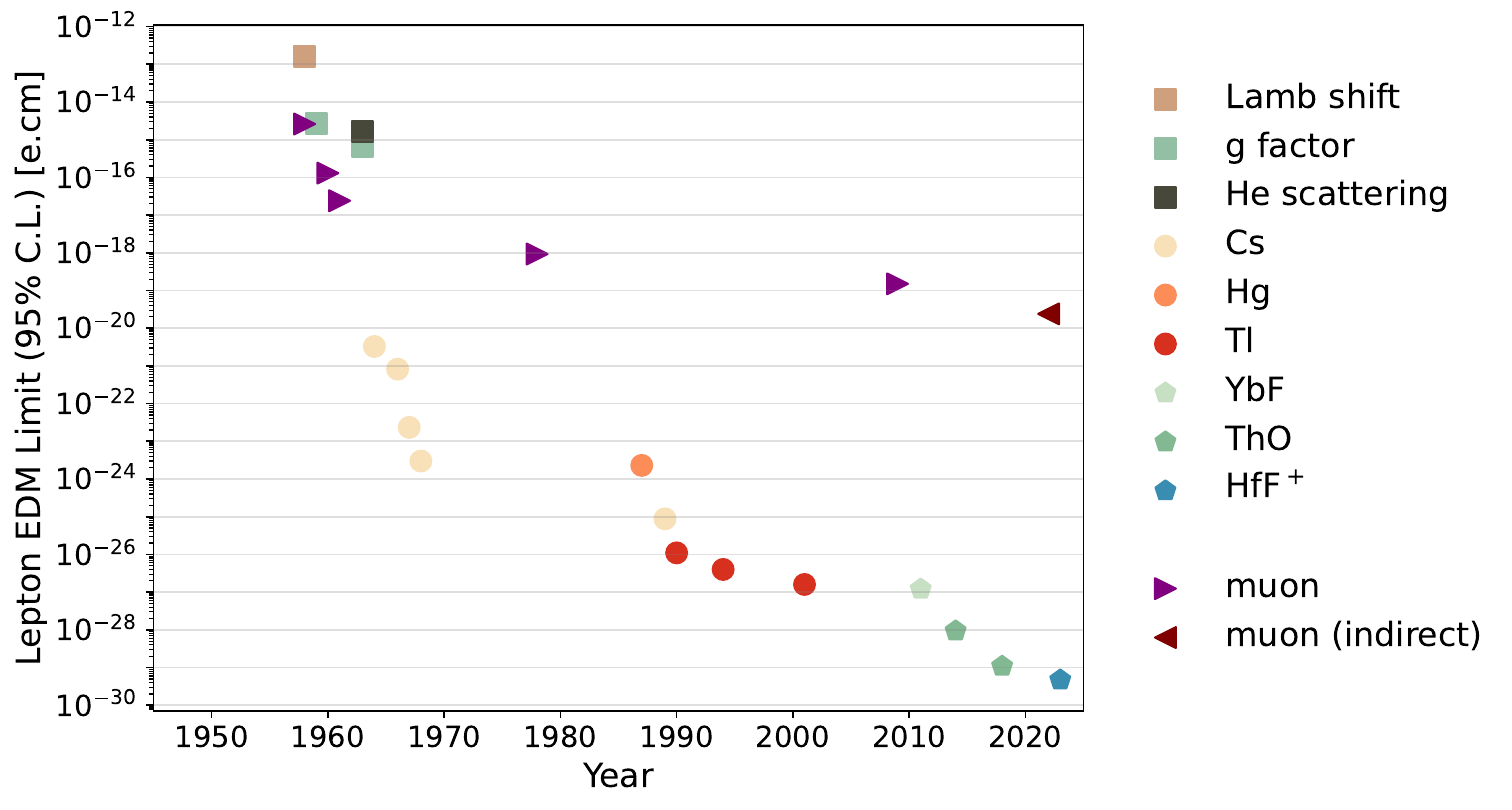}
\caption{\textbf{Electron and muon EDM limits}. The history of electron EDM limits, indirectly derived from various systems. The most sensitive measurements today use molecules, the best one on the HfF$^+$ ion setting a limit of $5\times10^{-30}e$cm for the electron~\cite{Roussy2023}. Opposite to the case of the diamagnetic $^{199}$Hg atom with its suppression factor (see Fig.\,\ref{fig:nucleonEDM}) the paramagnetic atoms and molecules exhibit large amplification factors. Also shown are the limits on the muon EDM, the best one published today still coming from the BNL muon $g$-2 experiment~\cite{Bennett2009}. Interestingly, the muon is the only measurement on a bare fundamental particle. Figure credit: T.~Hume, PSI.}
\label{fig:leptonEDM}
\end{figure}

\subsection{The story of ultracold neutrons (UCN) and exotic spin-dependent interactions}
\label{subsection:UCN}

Not long after Chadwick's 1932 discovery of the electrically-neutral neutron~\cite{Chadwick1932}, a spin-$1/2$ particle like the proton, it was shown that the neutron possessed a magnetic moment of a size similar to that of the proton. Along with the strong deviation of both the neutron and proton gyromagnetic ratios away from the value of 2 expected in Dirac's relativistic theory of the electron, this was a strong hint that neutrons and protons both probably possessed an internal structure of some type. Decades later, electron-scattering experiments finally resolved the quarks  whose dynamical motion and intrinsic magnetic moments combine to generate the magnetic properties of the neutron and proton, see, for example,\cite{Deur2019}. 
In the last few decades, the scientific mystery has shifted from the magnetic moments back to the spins of the neutron and proton: where exactly does the spin come from? The most obvious guess, namely that two of the three spins of the spin-$1/2$ quarks inside the proton and neutron simply cancel, is wrong experimentally. We now know that a large fraction of the proton and neutron spin comes from the electrically-neutral gluon field, which seems to confine the quarks inside protons and neutrons. The scientific lesson is clear: one can learn about fundamental questions by paying close attention to spin and the dynamical properties associated with it. 

Soon after neutron beams from nuclear reactors became available for study in the late 1940s, scientists were quick to take advantage of the special combination of properties possessed by these slow neutron beams.  Electrically neutral, but just magnetic enough for the spin to be manipulated delicately: able to penetrate macroscopic amounts of matter, while still interacting coherently. 

Fermi and Zinn~\cite{Fermi1946} showed that slow neutrons can experience total external reflection from flat surfaces, in contrast to the internal reflection of light in most media, thereby initiating a key technology for slow neutron manipulation. Zel'dovich later noted~\cite{Zeldovich1959} that one can confine very low energy neutrons, now called ``ultracold'' neutrons (UCN) inside material bottles. Shapiro~\cite{Shapiro1968} suggested the use of polarized UCN to search for the neutron EDM, and led one of the first experiments~\cite{Luschikov1969} which, in parallel with Steyerl's work~\cite{Steyerl1969}, successfully extracted and measured UCN for the first time. Now spin-polarized UCN ensembles are created nearly at rest in the lab to extend the measurement time for EDM and neutron beta decay measurements. UCN are created not by laser cooling as done for atoms (since there are no neutron-electron bound states), but by phonon cooling via neutron creation of phonons in a cold, dense medium from which the cooled neutrons can escape and be guided through tubes to the measurement chamber. 

\subsection{Why measure EDMs?}
\label{subsec:EDM searches}


As discussed in Secs.\,\ref{subsec:Dirac} and\,\ref{subsec:EDM history},  EDMs violate parity (P) and time reversal (T) symmetries.  If  physical reality is described by a quantum field theory, the CPT theorem (the C refers to charge conjugation) guarantees CPT invariance, based on most fundamental assumptions~\cite{Luders1957a, Luders1957b, pauli1955exclusion, Greaves2014, Blum2022}. The T-violating character of EDM therefore immediately corresponds to a CP violation, with CP being the symmetry connecting matter and antimatter. It turns out that the sensitive searches for finite permanent EDM present crucial tests of CP invariance, and the absence of signals so far provides some of the most constraining bounds on models and theories of CP violation `beyond the Standard Model' (SM, BSM) of particle physics. 
As the SM is believed to not provide sufficient amounts of CP violation to explain the observed baryon asymmetry of the universe~\cite{Sakharov1967,Riotto1999}, many such BSM models are motivated by the need to provide additional sources of CP violation. In general, many BSM theories naturally introduce CP violating parameters constrained only by experimental results. The known CP violation of the electroweak SM induces finite EDM, however, so far still many orders of magnitude smaller than experimental sensitivities. Also the strong interaction of the SM allows for a CP violating phase which is, however, already severely constrained by experimental limits on the neutron and nuclear EDM to be small of order of at most 10$^{-10}$ or zero, constituting the so-called `strong CP problem'. For an older but excellent overview, see, for example, \cite{Pospelov2005}. 

Over the years, the field of EDM searches has become much more diverse compared to the earlier work (Sec.\,\ref{subsec:EDM history}), 
and in particular, atoms and molecules entered the arena, allowing for extremely sensitive measurements of the electron and nuclear EDMs due to large possible amplification factors; see Fig.\,\ref{fig:leptonEDM}. A review of the present status of the field is beyond the scope of this article; however, we note that there is now a vibrant community and a plethora of approaches to searches for permanent electric dipole moments~\cite{Athanasakiskaklamanakis2026}, including initially unexpected directions, such as the use of polyatomic molecules and using storage rings as high-precision traps for charged particles. 



\section{Exotic spin-dependent interactions}
\subsection{Why exotic interactions?}

In the same marriage of relativity and quantum mechanics from which spin naturally emerges also comes the demand that interactions between particles be mediated locally through the exchange of a force-carrier. As for all particles, the force carrier must have mass (which can be zero, as for the photon and graviton) and spin, and different forces possess different amplitudes per unit spacetime volume (which we call charges or couplings), to emit and absorb the associated force carrier. The basic structure of the interactions in the non-gravitational component of our present Standard Model of particles and interactions emerges if one labels the force associated with one type of charge electromagnetic, the force associated with two types of charge that can transform into each other weak, and the force associated with three types of charge that can transform among each other strong.  Combined with the demand that the probability for any process in quantum mechanics cannot be greater than one, which leads to the conservation of the associated charge, most of the general features of the so-called strong, electromagnetic, and weak interactions emerge. Since the fundamental constituents of matter we know of are all spin 1/2 particles, the spin of the force carrier in electromagnetism, the photon, must be at least spin 1 (otherwise like charges in electrostatics would attract rather than repel), and likewise the W and Z bosons of the weak interaction and the gluons of the strong interaction are also spin 1. These interactions are implemented mathematically through the gauge field principle and result in renormalizable quantum field theories. Once again, we see that spin dictates many features of the forces we see in nature.


It is always possible that new weak forces await discovery. But what makes a force weak? It is not just the strength of the charge that matters: even if the charge is strong, if the spatial range of the force carrier is small due to its large mass, the force never gets anywhere. This leads to two different types of weak forces: one type has large charges and large mass for the force carrier, and the other has small charges and small mass for the force carrier. The first option attracted the most attention after the Standard Model was established as it is easier to make consistent with the idea of the unification of the three nongravitational forces into one. Possible new short-range interactions from new heavy particles are now classified using effective field theory techniques.   

With time and with the discovery of dark matter and dark energy the second option of new weakly-coupled longer-range interactions is being taken more seriously. Fortunately the infinite number of possibilities one might think this idea would lead to are still tightly constrained if one adheres to the same ideas of relativity and quantum mechanics which birth spin. For this reason there has been an explosion of experimental searches for weakly-coupled interactions mediated by light particle exchange, and weak spin-dependent interactions are natural candidates for forces that might have been missed in previous experiments.

\subsection{History of the exotic potentials} 

All of the known forces except gravity possess spin-dependent components to their interactions that lead to qualitatively distinct physical effects which have been clearly identified in experiments. However there is a vast unexplored range in coupling strength and exchange boson mass of possible exotic spin-dependent interactions which could exist, and a comparably wide variety of theoretical speculations which can produce exotic spin-dependent interactions as by-products of their attempts to explain various mysteries unaddressed in the standard model. This situation has motivated an explosion of experimental work to search for exotic spin-dependent effects over the last couple of decades.

Especially sensitive searches for such interactions can be conducted among the protons, neutrons, and electrons which compose normal matter, if one can design experiments and environmental conditions which suppress the background effects from the known interactions. Since all of the spin 1/2 particles in the standard model possess magnetic moments tied to the spin, any attempt to isolate an exotic spin-dependent interaction must at least deal with magnetic backgrounds. 

Since a large fraction of the experiments which have been conducted in this field employ matter in nonrelativistic motion in the lab, and since the effects are weak enough that they can be evaluated in the limit of single spin 0 or spin 1 boson exchange, it is convenient to classify the possibilities by taking the nonrelativistic limit of single boson exchange for all of the possible allowed types of couplings that can be present in the fermion currents (scalar, pseudoscalar, vector, axial vector, tensor) of the two interacting particles. This exercise generates a finite set of potentials which depend on the spins and velocities of the interacting particles. A recent review of this field~\cite{Cong2025_Spin-Dep_RMP} presents the current experimental constraints which come from a variety of experimental probes: torsion balances with polarized test masses and ensembles of optically-pumped atoms, which usually probe force ranges of macroscopic scale, NV centers in diamond and polarized slow neutron spin rotation, which probe mesoscopic distance scales, and analyses of precision measurements in atomic physics to prove Angstrom-scale interactions.   
    


The special properties of slow neutrons allow them to also probe even more fundamental aspects of spin. The wave function of a spin-1/2 particle in quantum mechanics is predicted to change sign under a $2\pi$ rotation: one must rotate the particle by $4\pi$ to return it back to the same state. One can test this prediction by passing a spin-1/2 particle into an interferometer, rotating the spin component of the quantum amplitude on one of the paths by an adjustable angle relative to that of the other path, and observing the interference pattern of the rotated and unrotated components of the state upon recombination. In the mid-1970s Rauch, Werner, and their collaborators developed a neutron interferometer using dynamical diffraction from cutout blades of a monolithic crystal of silicon. The capability of this device to separate the two paths of the neutron by macroscopic distances enabled precisely this experiment, with the relative rotation of the neutron spin on the two interferometer paths performed by magnetic fields which can precess the magnetic moment (and therefore the spin tied to it) by an adjustable angle. Their clear demonstration~\cite{Rauch1975, Werner1975} of the minus spin of the neutron spin under a $2\pi$ rotation is an example of one of the beautiful measurements testing and illustrating fundamental principles of quantum mechanics which this technology enabled. 

The minus sign of a spin-1/2 particle under a $2\pi$ rotation is consistent with the minus sign one gets in the exchange of two spin-1/2 particles from Fermi statistics, since the exchange of two identical particles is equivalent to a relative rotation of the two particles by $2\pi$. However, this minus sign is known to rely on the three-dimensionality of rotations in space: in cases where the system dynamics are constrained to two dimensions, one can in principle get any relative phase factor one wants under the interchange of two identical particles. The so-called ``anyons'', which live in such a two-dimensional world~\cite{Leinaas1977, Goldin1980, Wilczek1982}, can be realized in condensed matter and possess dynamics that make them interesting candidates for future applications in quantum information.

In 4D spacetime, the rotational invariance of space generalizes to 4D combinations of rotations and boosts. In fact, the CPT transformation can be viewed as a (complex) rotation in 4D spacetime~\cite{Jost1957}. The connection between spin and statistics~\cite{Fierz1939, Pauli1940} is not an independent ingredient in the CPT theorem: it follows already from two of the assumptions (energy positivity and microscopic causality) also needed in the CPT proof. What happens if CPT is violated? In this case, the spin-statistics connection can survive, but Lorentz symmetry must be violated~\cite{Greenberg2002}. This latter result forms the foundation for a wide variety of experimental searches for CPT/Lorentz violation, see, for example, review \cite{Safronova2018_RMP}.  

\subsection{Spin and gravity}


In view of the intimate connections between the properties of spin and the nature of spacetime, it is not surprising that speculations about possible connections between spin and spacetime geometry are a recurring theme in theoretical physics. Ever since the geometrization of gravity developed by Einstein which connected mass to spacetime curvature, there has long been a theoretical suspicion of a similar relationship of spin to spacetime geometry. Wigner's analysis of the quantum states of free point particles in relativistic spacetime in terms of irreducible unitary representations of the Poincare group~\cite{Wigner1939} concludes that all such particles can be classified by their mass and spin. It is also an interesting geometrical fact that, in 4D metric spacetime, there are two independent tensor structures characterizing the space that can appear, namely curvature and torsion. Is this match of the two properties of point particles and the two flavors of tensors in 4D curved spacetime geometry just a coincidence? If in addition one adds the locality requirement for symmetries and elevates the entire Poincare symmetry to a local symmetry, then one naturally gets a generalization of Einstein's theory called Einstein-Cartan gravity in which mass generates spacetime curvature and spin generates torsion~\cite{Cartan1922, Hehl1976}.  

Spin-dependent gravitational phenomena are already a part of standard general relativity through what one might think of as ``gravitomagnetic'' effects. Much like the magnetic moment dynamics of a particle moving in a magnetic field possesses contributions from both the orbital motion proportional to $\vec{L}$ and the spin contribution proportional to $\vec{s}$, the angular motion of particles in general relativity should also exhibit such effects. The legendary Gravity Probe B satellite resolved gravitomagnetic effects proportional to $\vec{L}$ in the curved spacetime of the earth, but so far no one has yet seen the associated spin effect, and the Gravity Probe Spin idea~\cite{Fadeev2021GPSpin} is named in its honor. 

\section{Spin, dark matter (DM), gravitational waves}
\label{Section: Spin, DM, GW}

The origin of dark matter (DM) apparently constituting some 80\% of all matter in the universe remains unknown. It may therefore be surprising that we can rather confidently state that if galactic dark matter consists primarily of particles of a certain kind, then, if the mass of the particle is less than a few electron-volts, the spin of this DM particle should be integer! But how can we know a spin property of an unknown particle? The argument goes like this. Suppose the particle has a half-integer spin. Then, according to the spin-statistics theorem, these particles are fermions. Now, the gravitational potential of a galaxy can be thought of as a ``box'' containing the DM particles, which assume a Fermi-Dirac distribution if the particles are fermions. For our galaxy, we know the local DM density to be about 0.4\,GeV/cm$^3$, while the escape velocity from the galaxy is on the order of $10^{-3}c$. A quick back-of-the-envelope estimate shows that for a gas of particles lighter than a few eV,
the Fermi velocity exceeds the galactic escape velocity, and because DM is associated with the galaxy, and does not ``fly away'', the particles cannot, in fact, be fermions and must instead be bosons, i.e., possess integer spin.

When it comes to the search for ultralight dark matter, spins play a central role as the core element of many detectors. This topic has been recently reviewed in \cite{aybas2026cavitylumpedcircuitspinbaseddetection}, with particularly sensitive detectors based on a mixture of alkali-metal and noble-gas spins discussed in \cite{Cong2023_Alcai_Noble_Gas}.
A recent realization is that modern spin-based sensors may also be sensitive to gravitational waves, see \cite{liang2025detectinggravitationalwavesspin} and references therein. 


\section{Big effects of a ``small'' spin}

The scale of elementary particle spin, $\hbar\approx 10^{-34}$\,J$\cdot$s, is minuscule compared to the angular momenta that we encounter in daily life. However, this in no way means that the effects of spin are ``small'' or even subtle. 

To start with, the entire structure of chemistry is governed by the fact that electrons are fermions. Should electrons have been bosons, the ground state of all atoms would have been $1s^Z$, where $Z$ is the atomic number, and there would have been nothing like a periodic table of elements. Incidentally, the ``boring'' bosonic atoms might, indeed, exist in nature. If there exist beyond-standard bosons like, for example, axions, they could accumulate around gravitating objects such as stars or planets just in such form \cite{Budker2023_Formation}.

We have already discussed in Sec.\,\ref{subsecProtonSpin} the example of parahydrogen, which shows that nuclear-spin properties govern the rotational states of dimer molecules, thus controlling energy scales that dramatically exceed those of direct spin-spin interaction.

An example at the interface of atomic and particle physics is positronium, the bound state of an electron and a positron. In this exotic form of ``hydrogen'', the ground state is $1S$, but the total spin can be either $J=0$ (parapositronium) or $J=1$ (orthopositronium), depending on how the spins of the electron and positron add together. Again, the difference in the spin arrangement has dramatic effects on the atom. For example, the lifetime of orthopositronium (142\,ns) is three orders of magnitude longer than that of parapositronium (125\,ps).

Spin is already a major factor in things that have not even been discovered or whose origin and composition are yet to be understood. An example is a model where galactic dark matter consists of one type of particle with a mass less than a few electron-volts that only interacts via gravitational forces \cite{Jackson_Kimball2023_UBDM_book}, as discussed in Sec.\,\ref{Section: Spin, DM, GW}.

The list of giant effects of ``small'' spin can be continued, for example, with neutron stars and supernovae, but we stop here, hopefully, having already made the important point. 


\section{Conclusions}

We hope that this brief article sufficiently illustrates the key role that spin has played in fundamental physics in the 100 or so years since its discovery and may play in the years to come.

Clearly, the limited scope has not allowed us to elaborate on many important subfields of spin physics. For example, in Sec.\,\ref{subsection:UCN}, we have discussed that confining (trapping) ultracold neutrons in ``bottles'' represents a key enabling technology for studying fundamental spin properties such as EDMs. But more generally, 
confining spins plays a key role across the spectrum of fundamental experiments and applications of spins. Examples include the use of Teflon-coated storage cells for polarized hydrogen in hydrogen masers \cite{Kleppner1965hydrogen}, paraffin-coated cells used in atomic magnetometers \cite{Seltzer2013_coatings}, laser-cooled and trapped atoms and molecules with spin, charged particles of matter and antimatter, and atomic and molecular ions with spin confined in Paul or Penning traps or storage rings, etc.
Spin is also central to precise and sensitive measurements; for example, magnetometry, which is often an indispensable part of fundamental experiments \cite{budker2013optical}.


As spin is clearly a cornerstone of modern physics, and because physics is fundamentally an experimental science, it is essential to subject the foundations of physics to unrelenting experimental scrutiny. This is true even without a specific theoretical basis for a possible violation of the laws. Examples of this include tests of spin-statistics connection, Pauli Exclusion Principle for fermions, Bose-Einstein statistics for bosons, as well as tests of CPT and Lorentz Invariance \cite{Curceanu2018_CERN_Courier,Tino_2001_Spin_Stat,Safronova2018_RMP}. 

We look forward to the essential role spin will continue to play in addressing the mysteries of our universe and probing the most fundamental principles of physics.

\clearpage 

%
\bibliography{100yosBib} 
\bibliographystyle{sciencemag}

%
%
%
%
%
%


\section*{Acknowledgments}
The authors are grateful to E.\,Willoughby, S.\,Ulmer, and T.\,Hume for help with the illustrations.  
The work of D.B. was supported by the ERC-2024-SYG 101167211 grant (GravNet), by the Cluster of Excellence ``Precision Physics, Fundamental Interactions, and Structure of Matter'' (PRISMA++ EXC 2118/2) funded by the German Research Foundation (DFG) within the German Excellence Strategy (Project ID 390831469), and by the COST Action within the project COSMIC WISPers (Grant No. CA21106). T.C. acknowledges support from the US National Science Foundation and the University of Michigan.
W. M. Snow acknowledges support from US National Science Foundation grant PHY-2209481, the US Department of Energy grant DE-SC0023695, and the Indiana University Center for Spacetime Symmetries.

\paragraph*{Competing interests:}
There are no competing interests to declare.
\paragraph*{Data and materials availability:}
All data and illustrations are either available in open access or from the sources indicted in the text.

\end{document}